\pdfoutput=1

\documentclass{iSWAGArticle}
\usepackage{todonotes}
\usepackage{float}
\usepackage{graphicx}
\usepackage{flushend}

\usepackage{amsmath}
\usepackage{amsthm}
\usepackage{hyperref}
\newcommand{\vbyte}{VByte}
\newcommand{\maskedvbyte}{\textsc{Masked \vbyte{}}}
\usepackage[utf8x]{inputenc}
\usepackage{booktabs}
\usepackage{listings}
\usepackage{algorithm}
\usepackage[noend]{algorithmic}
\usepackage[font=footnotesize]{subcaption}
\usepackage{tikz}
\usepackage{color}
\usetikzlibrary{chains,positioning,arrows}

\usepackage{times}

\usepackage{siunitx}

\graphicspath{{./results/}}
\usepackage{url}

\begin{document}
\title{Vectorized \vbyte{} Decoding} 

\author{\iSWAGAuthor{Jeff Plaisance\\
Indeed\\
jplaisance@indeed.com} \and \iSWAGAuthor{Nathan Kurz\\
Verse Communications\\
nate@verse.com} \and \iSWAGAuthor{Daniel Lemire\\
LICEF, Universit\'e du Qu\'ebec\\
lemire@gmail.com}}


\maketitle

\begin{abstract}
We consider the ubiquitous technique of \vbyte{} compression, which represents each integer as a variable length sequence of bytes. The
low 7 bits of each byte encode a portion of the integer, and the high bit of each byte is reserved as a continuation flag.  This flag is set to 1 for all bytes except the last, and the decoding of each integer
is complete when a byte with a high bit of 0 is encountered.  \vbyte{}
decoding can be a performance bottleneck especially when the
unpredictable lengths of the encoded integers cause frequent branch
mispredictions.  Previous attempts to accelerate \vbyte{} decoding using
SIMD vector instructions have been disappointing, prodding search
engines such as Google to use more complicated but faster-to-decode
formats for performance-critical code. Our decoder (\maskedvbyte{}) is 2
to 4~times faster than a conventional scalar \vbyte{} decoder, making the
format once again competitive with regard to speed.
\end{abstract}

\section{Introduction}
In many applications, sequences of integers are compressed with \vbyte{} to reduce memory usage.
For example, it is part of the search engine Apache Lucene (under the name vInt). It is used by Google in its Protocol Buffers interchange format (under the name Varint) and it is
part of the default API in the Go programming language.
It is also used in databases such as IBM DB2 (under the name Variable Byte)~\cite{Bhattacharjee:2009:EIC:1687553.1687573}.

We can describe the format as follows.
Given a non-negative integer in binary format, and starting from the least significant bits, we write it out using  seven bits in each byte, with the most significant bit of each byte set to 0 (for the last byte), or to 1 (in the preceding bytes).
In this manner, integers in  $[0,2^7)$ are coded using a single byte, integers in $[2^7, 2^{14})$ use two bytes and so on. See Table~\ref{table:illustration} for examples.

The \vbyte{} format is applicable to arbitrary integers including 32-bit and 64-bit integers. However, we focus on 32-bit integers for simplicity.


  \begin{table}[b]
  \caption{\label{table:illustration}\vbyte{} form for various powers of two. Within each word,  the most significant bits are presented first. In the \vbyte{} form, the most significant bit of each byte is in bold. }
  \centering\scriptsize
  \begin{tabular}{ccl} \toprule
  integer & binary form (16 bits) & \vbyte{} form \\ \midrule
  1 & 0000000000000001 &  \textbf{0}0000001\\
2 & 0000000000000010 & \textbf{0}0000010\\
4 & 0000000000000100 & \textbf{0}0000100\\
128 & 0000000010000000 & \textbf{1}0000000,  \textbf{0}0000001\\
256 & 0000000100000000 & \textbf{1}0000000,  \textbf{0}0000010 \\
512 & 0000001000000000 & \textbf{1}0000000,  \textbf{0}0000100 \\
16384 & 0100000000000000 & \textbf{1}0000000,  \textbf{1}0000000,  \textbf{0}0000001\\
32768 & 1000000000000000 & \textbf{1}0000000,  \textbf{1}0000000,  \textbf{0}0000010\\
\bottomrule
  \end{tabular}
  \end{table}

  \paragraph{Differential coding}
  A common application in information retrieval is to compress
  the list of document identifiers in an inverted index~\cite{williams1999compressing}. In such a case, we would not
  code directly the identifiers ($x_1, x_2, \ldots$), but
  rather their successive differences (e.g., $x_1-0, x_2-x_1, \ldots$), sometimes called deltas or gaps. If the document identifiers
 are provided in sorted order, then we might expect the gaps to
 be small and thus compressible using \vbyte{}. We refer to this approach as \emph{differential coding}. There are several possible approaches to differential coding. For example, if there are no repeated values, we can subtract one from each difference  ($x_1-0, x_2-x_1-1, x_3-x_2-1, \ldots$) or we can subtract blocks of integers for greater speed ($x_1,x_2,x_3,x_4,x_5-x_1,x_6-x_2,x_7-x_3,x_8-x_4,\ldots$). For simplicity, we only consider gaps defined as successive differences ($ x_2-x_1, \ldots$). In this instance, we need to compute a
 prefix sum over the gaps to recover the original values (i.e., $x_i = (x_i-x_{i-1})+x_{i-1}$).

\section{Efficient \vbyte{} Decoding}

One of the benefits of the \vbyte{} format is that we can write  an efficient decoder using just a few lines of code in almost any programming language. A typical decoder applies
Algorithm~\ref{alg:optimizedvbytedecoder}.
In this algorithm,
the function \texttt{readByte} provides  byte values in $[0,2^8)$ representing a number $x$  in the \vbyte{} format.

Processing each input byte requires only a few inexpensive operations (e.g., two additions, one shift, one mask). However, each byte also involves a branch. On a recent Intel processor (e.g., one using the Haswell microarchitecture), a single mispredicted branch can incur a cost of 15~cycles or more. When all integers are compressed down to one byte, mispredictions are rare and the performance is high.
However, when both one and two byte values occur in close proximity  the branch may become less predictable and performance may suffer.

For differential coding, we modify this algorithm
so that it decodes the gaps and computes the prefix sum. It suffices to keep track of the last value decoded and add it to the decoded gap.

  \begin{algorithm}
\begin{algorithmic}[1]
\STATE $y\leftarrow$ empty array of 32-bit integers
\WHILE{input bytes are available}
\STATE $b \leftarrow  \textrm{\texttt{readByte}}()$
\STATE \textbf{if} $b< 128$ \textbf{then} append $b$ to $y$ and \textbf{continue}
\STATE $c \leftarrow  b \bmod 2^7 $
\STATE $b \leftarrow  \textrm{\texttt{readByte}}()$
\STATE \textbf{if} $b< 128$ \textbf{then} append  $c + b \times 2^7$ to $y$ and \textbf{continue}
\STATE $c \leftarrow  c + (b \bmod 2^7) \times 2^7$
\STATE $b \leftarrow  \textrm{\texttt{readByte}}()$
\STATE \textbf{if} $b< 128$ \textbf{then} append  $c + b \times 2^{14}$ to $y$ and \textbf{continue}
\STATE $c \leftarrow  c + (b \bmod 2^7) \times 2^{14}$
\STATE $b \leftarrow  \textrm{\texttt{readByte}}()$
\STATE \textbf{if} $b< 128$ \textbf{then} append   $ c + b \times 2^{21}$ to $y$ and \textbf{continue}
\STATE $c \leftarrow  c + (b \bmod 2^7) \times 2^{21}$
\STATE $b \leftarrow  \textrm{\texttt{readByte}}()$
\STATE append $c +  b \times 2^{28}$ to $y$
\ENDWHILE
\STATE \textbf{return} $y$
\end{algorithmic}
    \caption{Conventional  \vbyte{} decoder. The \textbf{continue} instruction returns the execution to the main loop. The \texttt{readByte} function returns the next available input byte.}\label{alg:optimizedvbytedecoder}
  \end{algorithm}

\section{SIMD Instructions}
Intel processors provide SIMD instructions operating on 128-bit registers (called \emph{XMM registers}). These
registers can be considered as vectors of two 64-bit integers, vector2 of four 32-bit integers, vectors of eight 16-bit integers or vectors of sixteen 8-bit integers.

We review the main SIMD instructions we require in Table~\ref{ref:simdinstructions}. We can roughly judge the computational cost of an instruction by its latency and reciprocal throughput. The latency is the minimum number of cycles required to execute the instruction. The latency is most important when subsequent operations have to wait for the instruction to complete. The reciprocal throughput is the inverse of the maximum number of instructions that can be executed per cycle. For example, a reciprocal throughput of 0.5 means that up to two instructions can be executed  per cycle.

We use the \texttt{movdqu} instruction to load or store a register. Loading and storing registers has a relatively high latency (3~cycles). While we can load two registers per cycle, we can only store one of them to memory.
A typical SIMD instruction is \texttt{paddd}: it adds two vectors of four 32-bit integers at once.

Sometimes it is necessary to selectively copy the content from one XMM register to another while possibly copying and duplicating components to other locations. We can do so with the \texttt{pshufd} instruction when considering the registers as  vectors of 32-bit integers, or
with the \texttt{pshufb} instruction when  registers is considered vectors of bytes. These instructions take an input register $v$ as well as a control mask $m$ and they  output a new vector $(v_{m_0},v_{m_1},v_{m_2},v_{m_3}, \ldots)$ with the added convention that $v_{-1}\equiv 0$. Thus, for example, the  \texttt{pshufd} instruction can copy one particular value to all positions (using a mask made of 4~identical values).  If we wish to shift by a number of bytes, it can be more efficient to use a dedicated instruction (\texttt{psrldq} or \texttt{pslldq}) even though the \texttt{pshufb} instruction could achieve the same result. Similarly, we can use the \texttt{pmovsxbd} instruction to more efficiently unpack the first four bytes as four 32-bit integers.

We can simultaneously shift right by a given number of bits all of the components of a vector using the instructions
\texttt{psrlw} (16-bit integers),
\texttt{psrld} (32-bit integers)
and \texttt{psrlq} (64-bit integers). There are also corresponding left-shift instructions such as \texttt{psllq}.
We can also compute the bitwise OR and bitwise AND between two 128-bit registers using the \texttt{por} and \texttt{pand} instructions.

There is no instruction to shift a vector of 16-bit integers by
different number of bits (e.g., $(v_1, v_2, \ldots)\to (v_1 \ll 1, v_2 \ll 2, \ldots)$)  but we can get the equivalent result by mutiplying integers (e.g., with the
\texttt{pmullw} instruction). The AVX2 instruction set introduced such  flexible shift instructions (e.g., \texttt{vpsrlvd}), and they are much faster than a multiplication, but they not applicable to vectors of 16-bit integers.  Intel proposed a new instruction set (AVX-512) which contains such an instruction (\texttt{vpsrlvw}) but it is not yet publicly available.

Our contribution depends crucially on the \texttt{pmovmskb} instruction. Given a vector of sixteen bytes, it outputs a 16-bit value made of the most significant bit of each of the sixteen input bytes: e.g., given the vector $(128, 128, \ldots, 128)$, \texttt{pmovmskb} would output 0xFFFF.

\begin{table*}[bth]
\caption{Relevant SIMD instructions on  \emph{Haswell} Intel processors with latencies and reciprocal throughput in CPU cycles \label{ref:simdinstructions}.}\centering\small
\begin{tabular}{cp{3.1in}cc}
\toprule
instruction & description & latency &  \multicolumn{1}{p{0.7in}}{rec.\ throughput}
\\\midrule
\texttt{movdqu} & store or retrieve a 128-bit register & 3 & 1/0.5\\
\texttt{paddd} & add four pairs of 32-bit integers& 1 & 0.5\\
\texttt{pshufd} & \emph{shuffle} four 32-bit integers& 1 & 1\\
\texttt{pshufb} & \emph{shuffle}  sixteen bytes & 1 & 1\\
\texttt{psrldq} & shift right by a number of bytes & 1 & 0.5\\
\texttt{pslldq}& shift left by a number of bytes & 1 & 0.5\\
\texttt{pmovsxbd} & unpack the first four bytes into four 32-bit ints. & 1 & 0.5\\
\texttt{pmovsxwd} & unpack the first four 16-bit integers into four 32-bit ints. & 1 & 0.5\\
\texttt{psrlw} & shift right eight 16-bit integers & 1 & 1\\
\texttt{psrld} & shift right four 32-bit integers & 1 & 1 \\
\texttt{psrlq} & shift right two 64-bit integers & 1 & 1 \\
\texttt{psllq} & shift left two 64-bit integers & 1 & 1 \\
\texttt{por} & bitwise OR between two 128-bit registers & 1 & 0.33\\
\texttt{pand} & bitwise AND between two 128-bit registers  & 1 & 0.33\\
\texttt{pmullw} & multiply eight 16-bit integers & 5 & 1\\
\texttt{pmovmskb} & create a 16-bit mask from the most significant bits & 3 & 1\\
\bottomrule
\end{tabular}
\end{table*}

\section{\maskedvbyte{} decoding}

The conventional \vbyte{} decoders algorithmically process one input byte at a time (see Algorithm~\ref{alg:optimizedvbytedecoder}). To multiply the decoding speed, we want to process larger chunks of input data at once.
Thankfully, commodity Intel and AMD processors have supported \emph{Single instruction, multiple data} (SIMD) instructions since the introduction of the Pentium~4 in 2001. These instructions can process several words at once, enabling \emph{vectorized} algorithms.

Stepanov et al.~\cite{Stepanov:2011:SDP:2063576.2063627} used
SIMD instructions to accelerate the decoding of \vbyte{} data (which they call varint-SU). According to their experimental results, SIMD instructions lead to a disappointing speed improvement of less than \SI{25}{\percent}, with no gain at all in some instances. To get higher speeds (e.g., an increase of $3\times$), they proposed instead new formats akin to Google's Group Varint~\cite{DeanOfficialplusslides:2009:CBL:1498759.1498761}. For simplicity, we do not consider such ``Group'' alternatives further:  once we consider different data format, a wide range of fast SIMD-based compression schemes become available~\cite{LemireBoytsov2013decoding}---some of them faster than Stepanov et al.'s fastest proposal.

\begin{figure*}
\centering
\parbox{0.7\textwidth}{%
\begin{tikzpicture}[thick,scale=0.8, every node/.style={scale=0.8},node distance=0cm,start chain=9 going below,start chain=13 going right,start chain=14 going right,start chain=15 going right,start chain=16 going right,start chain=17 going right,start chain=18 going right,start chain=19 going right,start chain=20 going right,start chain=21 going right]
\edef\sizetape{0.07cm}    \tikzstyle{mytape}=[draw,minimum height=0.7cm,minimum width=0.5cm]
\tikzstyle{overbrace style}=[decorate,decoration={brace,raise=2mm,amplitude=3pt}]
    \node(B1) [on chain=9,mytape] {\textbf{1}0000000};
    \node(A3) [on chain=9,mytape] {\textbf{0}0000001};
    \node(A1) [on chain=9,mytape] {\textbf{1}0000010};
    \node(A2) [on chain=9,mytape] {\textbf{0}0000011};
    \node [on chain=9,mytape] {\textbf{0}0010000};
    \node [on chain=9,mytape] {\textbf{0}0100000};

                 \node [above of=B1,node distance=0.8cm] {Compressed data  (6~bytes)};

                     \node(E1) [on chain=13,mytape,right=2.5cm of B1] {1};
    \node [on chain=13,mytape] {0};
    \node [on chain=13,mytape] {1};
    \node(Z1) [on chain=13,mytape] {0};
    \node [on chain=13,mytape] {0};
    \node [on chain=13,mytape] {0};

            \node(M1) [above of=Z1,node distance=0.8cm] {Mask (6 bits)};
\draw[<->, >=latex', shorten >=2pt, shorten <=2pt, thick,dashed]
    (A3.east) to node[fill=blue!10] {\parbox{1.6cm}{\footnotesize Extract mask (\texttt{pmovmskb})}}
    (E1.west);

    \node(Z2) [on chain=14,mytape,right=2.5cm of A1] {\textbf{1}0000000};
    \node(ZP2) [on chain=14,mytape] {\textbf{0}0000001};

    \node(Z3) [on chain=15,mytape,below=0cm of Z2] {\textbf{1}0000010};
    \node [on chain=15,mytape] {\textbf{0}0000011};

    \node(Z4) [on chain=16,mytape,below=0cm of Z3] {\textbf{0}0010000};
    \node [on chain=16,mytape] {\phantom{000}0\phantom{0000}};

    \node(Z5) [on chain=17,mytape,below=0cm of Z4] {\textbf{0}0100000};
    \node(Z52) [on chain=17,mytape] {\phantom{000}0\phantom{0000}};

\draw[<->, >=latex', shorten >=2pt, shorten <=2pt, thick,dashed]
    (A2.east) to node[fill=blue!10] {\parbox{1.6cm}{\footnotesize Permute (\texttt{pshufb}) using mask and look-up table}}
    (Z3.west);
 \node [below of=Z1,node distance=0.8cm] {\parbox{3cm}{Permuted bytes}};

 \node [below of=Z5,node distance=0.8cm,fill=red!10] {\small low bytes};
 \node [below of=Z52,node distance=0.8cm,fill=red!10] {\small high bytes};

    \node(ZZ2) [on chain=18,mytape,right=2.5cm of Z2] {10000000};
    \node(ZZP2) [on chain=18,mytape] {\phantom{000}0\phantom{0000}};

    \node(ZZ3) [on chain=19,mytape,below=0cm of ZZ2] {10000010};
    \node [on chain=19,mytape] {00000001};

    \node(ZZ4) [on chain=20,mytape,below=0cm of ZZ3] {00010000};
    \node [on chain=20,mytape] {\phantom{000}0\phantom{0000}};

    \node(ZZ5) [on chain=21,mytape,below=0cm of ZZ4] {00100000};
    \node(ZZ52) [on chain=21,mytape] {\phantom{000}0\phantom{0000}};
            \node(BIG1) [right of=Z1,node distance=5cm,fill=blue!10] {\parbox{5cm}{\footnotesize Decoded values
            from the permuted bytes by extracting the 4~low and 4~high bytes with masks, shifting down by 1 bit the high bytes and
            ORing.}};
 \node [below of=BIG1,node distance=4.5cm] {\small Integers 128, 386, 16, 32};

(128, 386, 16, 32)

\end{tikzpicture}
}
\caption{Simplified illustration of vectorized \vbyte{} decoding from 6~bytes to four 16-bit integers  (128, 386, 16, 32).  \label{fig:illustration}}
\end{figure*}
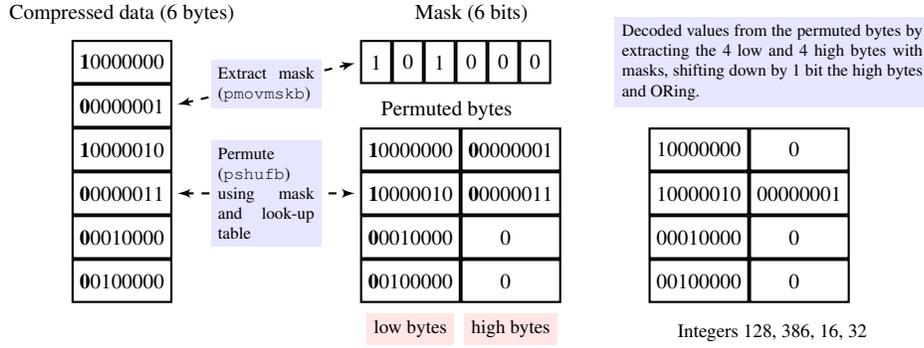

Though they did not provide a detailed description, Stepanov et al.'s approach resembles ours in spirit.
Consider the simplified example from Fig.~\ref{fig:illustration}.  It illustrates the main steps:
\begin{itemize}
\item From the input bytes, we gather the control bits (1,0,1,0,0,0 in this case) using the \texttt{pmovmskb} instruction.
\item From the resulting mask, we look up a \emph{control mask} in a table and apply the  \texttt{pshufb} instruction to move the bytes. In our example, the first 5~bytes are left in place (at positions 1, 2, 3, 4, 5) whereas the 5$^{\mathrm{th}}$ byte is moved to position 7. Other output bytes are set to zero.
\item We can then extract the first 7~bits of the low bytes (at positions 1, 3, 5, 7) into a new 8-byte register. We can also extract the high bytes (positions 2, 4, 6, 8) into another 8-byte register. On this second register, we apply a right shift by 1~bit on the four~16-bit values (using \texttt{psrlw}). Finally, we compute the bitwise OR of these two registers, combining the results from the low and high bits.
\end{itemize}

A na\"ive implementation of this idea could be slow. Indeed, we face several performance challenges:
\begin{itemize}
\item The \texttt{pmovmskb} instruction has a relatively high latency (e.g., 3~cycles on the Haswell microarchitecture).
\item The \texttt{pmovmskb} instruction processes 16~bytes at once, generating a 16-bit result. Yet looking up a 16-bit value in a table would require a 65536-value table. Such a large table is likely to stress the CPU cache.

Moreover, we do not know ahead of time where coded integers begin and end: a typical segment of 16~bytes might contain the end of one compressed integer, a few compressed integer at the beginning of another compressed integer.
\end{itemize}

Our proposed algorithm works on 12~bytes inputs and 12-bit masks.
In practice, we load the input bytes in  a 128-bit register
containing 16~bytes (using \texttt{movdqu}), but only the first 12~bytes are considered.
 For the time being, let us assume that the segment begins with a complete encoded integer.   Moreover, assume that the 12-bit mask has been precomputed.

In what follows, we use the convention that $(\cdots)_k$ is a vector of $k$-bit integers. Because numbers are stored in binary notation, we have that  \begin{align*}(1,0,0,0)_8 = (1,0)_{16} = (1)_{32},\end{align*} that is, all three vectors represent the same binary data.

\begin{itemize}
\item If the mask is $00\cdots 00$, then the 12~input bytes represent 12~integers \emph{as is}. We can unpack the first
4~bytes to 4~32-bit integers (in a 128-bit register) with the \texttt{pmovsxbd} instruction. This new register can
then be stored in the output buffer. We can then shift the input register by 4~bytes using the \texttt{psrldq} instruction, and apply to \texttt{pmovsxbd} instruction again. Repeating a third time, we have decoded all 12~integers.
We have consumed 12~input bytes and written 12~integers.
\item Otherwise we use the 12-bit mask to look two 8-bit values in
a table $2^{12}$-entries-wide. The
first 8-bit value is an integer between 2 and 12 indicating
how many input bytes we consume.
Though it is not immediately
useful to know how many bytes are consumed, we use this number of consumed bytes when loading the next input bytes. The second 8-bit value is an \emph{index} $i$ taking integer values in $[0,170)$. From this index, we load up one of 170~control masks. We then proceed according to the value of the index $i$:
\begin{itemize}
\item If $i<64$, then  the next 6~integers each fit in at most two bytes (they are less than $2^{14}$). There are exactly $2^6=64$~cases corresponding to this scenario. That is, the first integer can fit in one or two bytes, the second integer in one or two bytes, and so on, generating 64~distinct cases. For each of the 6~integers $x_i$, we have the low byte containing the least significant 7~bits of the integer $a_i$, and optionally a high byte containing the next 7~bits $b_i$ ($x_i= a_i + b_i 2^{7}$).
The call to \texttt{pshufb} will
permute the bytes such that the low bytes occupy the positions 1, 3, 5, 7, 9, 11 whereas the high bytes, when available, occupy the positions 2, 4, 6, 8, 10, 12. When a high byte is not available, the byte value zero is used instead.

For example, when all 6~values are in $[2^{7},2^{14})$, the permuted bytes are \begin{align*}(a_1 1,b_1,a_2 1,b_2,a_3 1,b_3,\ldots,a_6 1,b_6)_8\end{align*} when presented as a vector of bytes with the short-hand notation $a_i 1 \equiv a_i + 2^{7}$.

From these permuted bytes, we generate two vectors using bitwise ANDs with fixed masks (using \texttt{pand}). The first one retains only the least significant 7~bits of the low bytes: as a vector of 16-bit integers we have \begin{align*}
& (a_1 1,b_1,a_2 1,b_2,a_3 1,b_3,\ldots,a_6 1,b_6)_8
\\ &\textrm{~AND~} \\
& (127,0,127,0,127,0,\ldots,127,0)_8 \\
& =(a_1, a_2,a_3,\ldots,a_6)_{16}.\end{align*} The second one retains only the high bytes: \begin{align*}(0,b_1,0,b_2,0,b_3,\ldots,0,b_6)_8.\end{align*}
Considering the latter as a vector of 16-bit integers, we right shift it by 1~bit (using \texttt{psrlw}) to get the following vector \begin{align*}(b_1   2^7,b_2  2^7,b_3  2^7,\ldots,b_6  2^7)_{16}.\end{align*}  We can then combine (with a bitwise OR using \texttt{por}) this last vector with the vector containing the least significant 7~bits of the low bytes. We have effectively decoded the 6~integers as 16-bit integers: we get
\begin{align*}(a_1 + b_1    2^7, a_2 + b_2    2^7,\\
a_3 + b_3    2^7, a_4 + b_4    2^7,\\
a_5 + b_5    2^7, a_6 + b_6    2^7)_{16}.\end{align*}
 We can unpack the first four to 32-bit integers using an instruction such as \texttt{pmovsxwd}, we can then shift by 8~bytes (using \texttt{psrldq}) and apply \texttt{pmovsxwd} once more to decode the last two integers.

\item If $64 \leq i <145$,  the next 4~encoded integers fit in at most 3~bytes. We can check that there are $81=3^4$~such cases.
The processing is then similar to the previous case except that
we have up to three bytes per integer (low, middle and high).
The permuted version will re-arrange the input bytes so that the
first 3~bytes contain the low, middle and high bytes of the first integer, with the convention that a zero byte is written when there is no corresponding input byte. The next byte always contain a zero. Then we store the data corresponding to the next integer in the next 3 bytes. A zero byte is added. And so on.

This time, we create 3~new vectors using bitwise ANDs with appropriate masks: one retaining only the least significant 7~bits from the low bytes, another retaining only the least significant 7~bits from the middle bytes and another retaining only the high bytes. As vectors of 32-bit integers, the second vector is right shifted by 1~bit  whereas the third vector is right shifted by 2~bits (using \texttt{psrld}). The 3~registers are then combined with a bitwise OR and written to the output buffer.

\item Finally, when $145\leq i < 170$), we decode the next 2~integers. Each of these integers can consume from 1~to
5~input bytes. There are $5^2=25$ such cases.

For simplicity of exposition, we only explain how we decode the first of the two integers using 8-byte buffers. The integer can be written as $x_1 = a_1 +  b_1 2^7+  c_1 2^{14} +  d_1 2^{21}+  e_1 2^{28}$ where $a_1,b_1,c_1,d_1 \in [0,2^7)$ and $e_1 \in [0,2^4)$. Assuming that $x_1 \geq 2^{28}$, then the first 5~input bytes will be $(a_1 1, b_1 1, c_1 1, d_1 1, e_1)_8$.

Irrespective of the value of the index $i$, the first step is to set the most significant bit of each input byte to 0 with a bitwise AND\@.  Thus, if $x_1\geq 2^{28}$, we get $(a_1, b_1, c_1, d_1, e_1)_8$.

We then permute the bytes so that we get the following 8~bytes: \begin{align*}Y=( b_1,  c_1, d_1, e_1 + a_1 2^8)_{16}.\end{align*}
The last byte is occupied by the value $a_1$ which we can isolate for later use by
shifting right the whole vector by seven bytes (using \texttt{psrldq}):
\begin{align*}Y'=(a_1,0,0,0,0,0,0,0)_{8}.\end{align*}
Using the \texttt{pmullw} instruction, we multiply the permuted bytes ($Y$) by the vector
\begin{align*}(2^{7},  2^{6}, 2^{5}, 2^{4})_{16}\end{align*}
to get
\begin{align*}( b_1 2^{7},  c_1 2^{6}, d_1 2^{5}, e_1 2^{4} + (a_1 2^{12} \bmod 2^{16}))_{16}.\end{align*}
As a byte vector, this last vector is equivalent to
\begin{align*}X=( b_1 2^7 \bmod 2^8,  b_1 \div 2,
\\ c_1 2^6 \bmod 2^8
, c_1 \div 2^2,\\
d_1 2^5 \bmod 2^8,
d_1 \div 2^3,\\
e_1 2^6 \bmod 2^8,
\star)_{8}\end{align*}
where we used $\star$ to indicate an irrelevant byte value.
We can left shift this last result by one byte (using \texttt{psllq}):
\begin{align*}X'=(0, b_1 2^7 \bmod 2^8,\\  b_1 \div 2, c_1 2^6 \bmod 2^8
,\\ c_1 \div 2^2,
d_1 2^5 \bmod 2^8,\\
d_1 \div 2^3,
e_1 2^6 \bmod 2^8)_{8}\end{align*}
We can combine these two results with the value $a_1$ isolated earlier ($Y'$):
\begin{align*} Y'\mathrm{~OR~}X\mathrm{~OR~}X=(a_1 + b_1 2^7 \bmod 2^8,\star,\\
 b_1 \div 2 +  c_1 2^6 \bmod 2^8, \star
,\\ c_1 \div 2^2 +d_1 2^5 \bmod 2^8,
\star,\\
d_1 \div 2^3 + e_1 2^6 \bmod 2^8,
\star)_{8}\end{align*}
where again we use $\star$ to indicate  irrelevant byte values.
We can  permute this last vector to get
\begin{align*}& (a_1 + b_1 2^7 \bmod 2^8, b_1 \div 2 +  c_1 2^6 \bmod 2^8,\\& c_1 \div 2^2 +d_1 2^5 \bmod 2^8,\\
&d_1 \div 2^3 + e_1 2^6 \bmod 2^8,
\ldots)_{8}\\
&= (a_1 +  b_1 2^7+  c_1 2^{14} +  d_1 2^{21}+  e_1 2^{28},\ldots)_{32}\end{align*}
Thus, we have effectively decoded the integer $x_1$.

The actual routine works with two integers $(x_1, x_2)$. The content of the first one is initially stored in the first 8~bytes of a 16-byte vector whereas the remaining 8~bytes are used for the second integer. Both integers are decoded simultaneously.
\end{itemize}

\end{itemize}

An important motivation is to amortize the latency of the \texttt{pmovmskb} instruction  as much as possible.
 First, we repeatedly call the \texttt{pmovmskb} instruction until we have processed up to 48~bytes to compute a corresponding 48-bit mask. Then we repeatedly call the decoding procedure as long as 12~input bits remain out of the processed 48~bytes. After reach call to the 12-byte decoding procedure, we  left shift the mask by the number of consumed bits. Recall that we look up the number of consumed bytes at the beginning of the decoding procedure so this number is readily available and its determination does not cause any delay. When fewer than 12~valid bits remain in the mask, we process another block of 48~input bytes with \texttt{pmovmskb}.
To accelerate further this process, and if there are enough input bytes, we maintain two 48-bit masks (representing 96~input bytes): in this manner, a 48-bit mask is already available while a new one is being computed.
 When fewer than 48~input bytes but more than 16~input bytes remain, we call \texttt{pmovmskb} as needed to ensure that we have at least a 12-bit mask. When it is no longer possible, we fall back on conventional \vbyte{} decoding.

\paragraph{Differential Coding}

We described the decoding procedure without accounting for
differential coding. It can be added without any major
algorithmic change. We just keep track of last
32-bit integer decoded. We might store it in the last entry of a vector of four 32-bit integers (henceforth $p=(p_1,p_2,p_3,p_4)$).

We compute the prefix sum before to writing the decoded integers.
There are two cases to consider. We either write four 32-bit integers or 2 32-bit integers (e.g., when writing 6~decoded integers, we first write 4, then 2 integers). In both cases, we permute first the entries of $p$ so that $p\leftarrow (p_4,p_4,p_4,p_4)$ using the \texttt{pshufd} instruction.
\begin{itemize}
\item Suppose we decoded four integers in the vector $c$. We left shift the content of $c$ by one integer (using \texttt{pslldq}) so that $c'\leftarrow (0,c_1,c_2,c_3)$. We add $c$ to $c'$ using \texttt{paddd} so that $c\leftarrow (c_1,c_1+c_2,c_2+c_3,c_3+c_4)$. We shift the rest by two integers (using again \texttt{pslldq}) $c'\leftarrow (0,0,c_1,c_1+c_2)$ and add $c+c' =(c_1,c_1+c_2,c_1+c_2+c_3,c_1+c_2+c_3+c_4)$. Finally, we add $p$ to this last result $p\leftarrow p + c+c'=(p_4+c_1,pp+c_1+c_2,p+c_1+c_2+c_3,p+c_1+c_2+c_3+c_4)$. We can write $p$ as the decoded output.
\item The process is similar though less efficient if we only have two decoded gaps. We start from a vector containing two gaps $c\leftarrow (c_1,c_2,\star, \star)$ where we indicate irrelevant entries with $\star$. We can left shift by one integer $c'\leftarrow (0,c_1,c_2,\star)$ and add the result $c+c'=(c_1,c_1+c_2,\star,\star)$. Using the \texttt{pshufd} instruction, we can copy the value of the second component to the third and fourth components, generating
$(c_1,c_1+c_2,c_1+c_2,c_1+c_2)$. We can then  add $p$ to this result and store the result back into $p$. The first two integers can be written out as output.
\end{itemize}

%

\section{Experiments}

We implemented our software in C and  C++.
The benchmark program ran on  a Linux server with an Intel i7-4770 processor running at \SI{3.4}{GHz}.
This Haswell processor has \SI{32}{kB} of L1 cache and \SI{256}{kB} of L2 cache per core with \SI{8}{MB} of L3 cache.
The machine has \SI{32}{GB} of RAM (DDR3-1600 with double-channel). We disabled Turbo Boost and set the processor
to run at its highest clock speed. We report wall-clock timings.
Our software is freely available
under an open-source license (\url{http://maskedvbyte.org})
and was  compiled using the GNU GCC~4.8 compiler with  the \texttt{-O3} flag.

\begin{figure}[t]\centering
    \begin{subfigure}{.49\textwidth}
\includegraphics[width=1\textwidth]{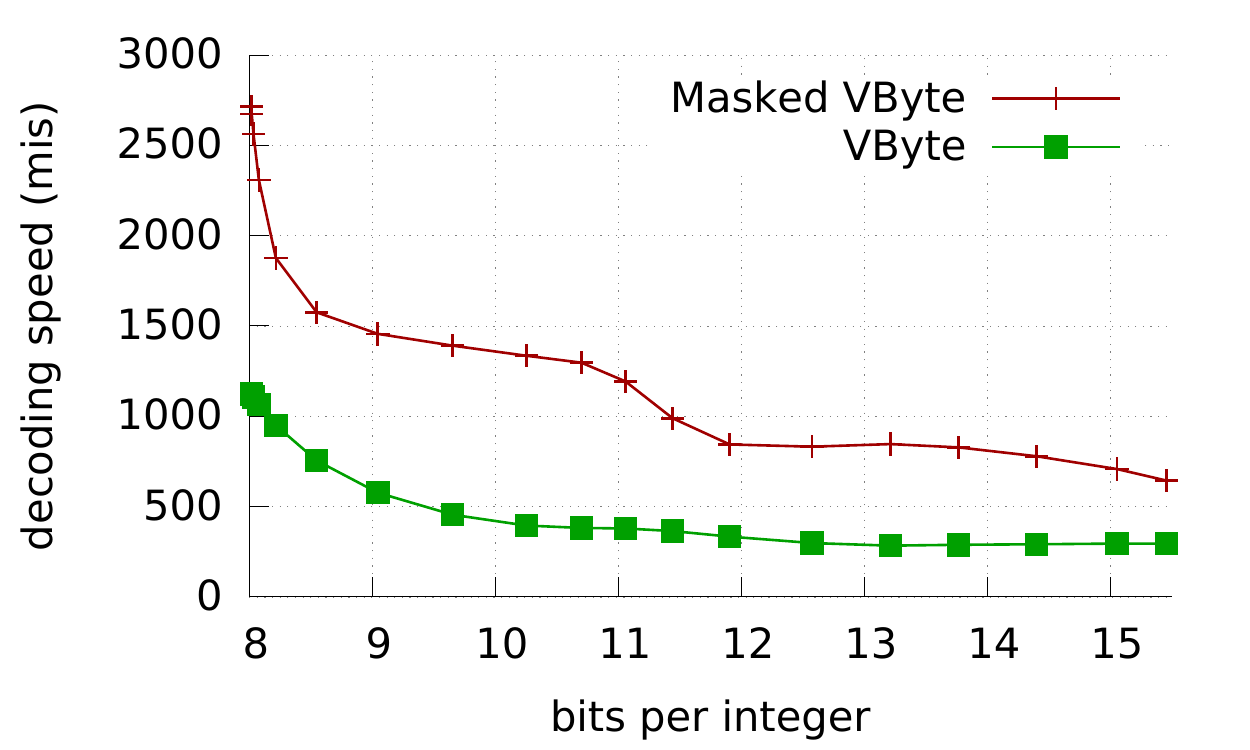}
\subcaption{Absolute speed}
\end{subfigure}
    \begin{subfigure}{.49\textwidth}
\includegraphics[width=1\textwidth]{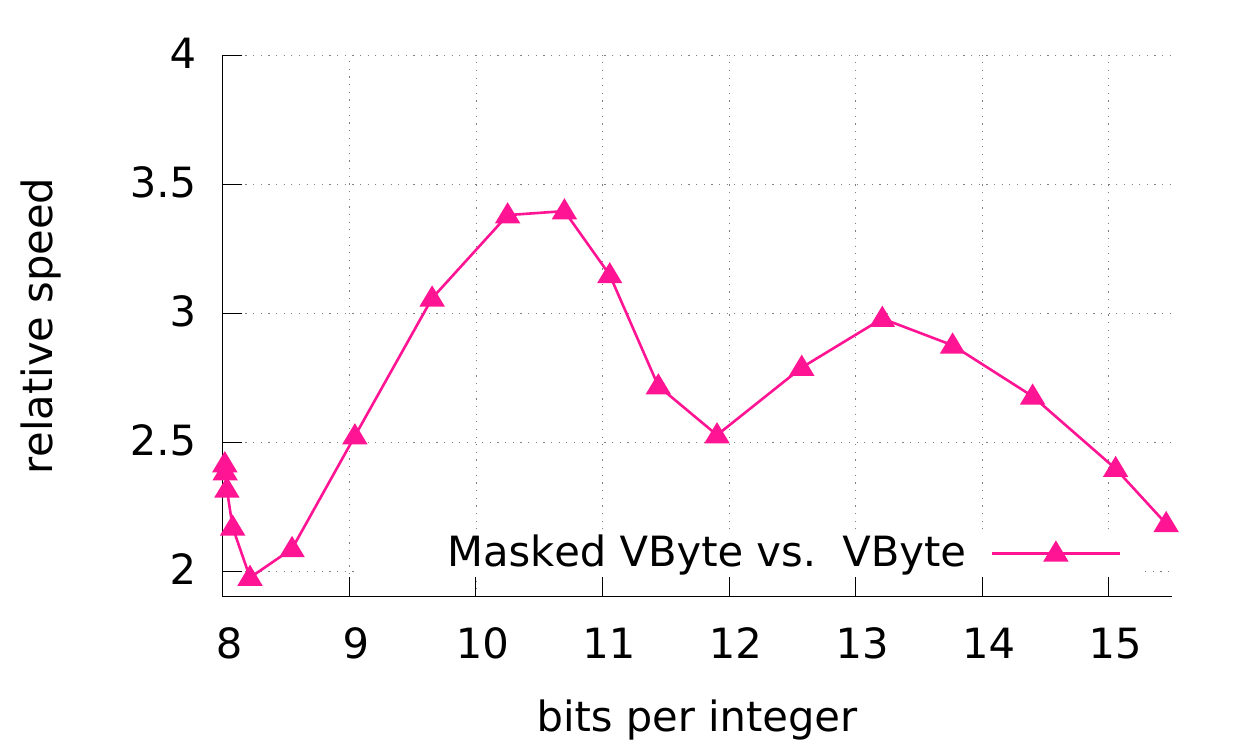}
\subcaption{Ratio of speeds}
\end{subfigure}
\caption{\label{fig:comp}Performance comparison for various sets of posting lists (ClueWeb)}
\end{figure}

For our experiments, we used a collection of posting lists extracted from the ClueWeb09 (Category B) data set. ClueWeb09 includes 50~million web pages. We have one posting list for each of the 1~million most frequent words---after excluding stop words and applying lemmatization. Documents were sorted lexicographically based on their URL prior to attributing document identifiers.
The posting lists are grouped based on  length:
we store and process lists of lengths  $2^K$ to $2^{K+1}-1$ together for all values of $K$.  Coding and decoding times include differential coding. Shorter lists are less compressible  than longer lists since their gaps tend to be larger.
 Our results are summarized in Fig.~\ref{fig:comp}. For each group of posting lists we compute the average bits used per integer after compression: this value ranges from 8 to slightly less than 16. All decoders work on the same compressed data.

  When decoding long posting lists to RAM, our speed is limited by RAM throughput. For this reason,
 we decode the compressed data sequentially to buffers fitting in L1 cache (4096~integers).
 For each group and each decoder, we compute the average decoding speed in millions of 32-bit integers per second (mis). For our \maskedvbyte{} decoder, the speeds ranges from \SI{2700}{mis} for the most compressible lists to \SI{650}{mis} for the less compressible ones. The speed of the conventional\vbyte{} decoder ranges from \SI{1100}{mis} to \SI{300}{mis}. For all groups of posting lists in our experiments, the \maskedvbyte{} decoder was at least twice as fast as the conventional \vbyte{} decoder. However, for some groups, the speedup is between  $3\times$ and $4\times$.

If we fully decode all lists instead of decoding to a buffer that fits in CPU cache,
the performance of \maskedvbyte{} can be reduced by about \SI{15}{\percent}. For example, instead of a maximal speed of \SI{2700}{mis},  \maskedvbyte{} is limited to \SI{2300}{mis}.

\section{Conclusion}

To our knowledge, no existing \vbyte{} decoder comes close to the speed of \maskedvbyte{}.  Given how the \vbyte{} format is a de facto standard, it suggests that \maskedvbyte{} could help optimize a wide range of existing software without affecting the data formats.

 \maskedvbyte{} is in production code at Indeed as part of the open-source analytics platform Imhotep (\url{http://indeedeng.github.io/imhotep/}).

\section*{Acknowledgments}

We thank L. Boystov from CMU for preparing and making available the posting list collection.

\bibliographystyle{plain}
\bibliography{varint}

\begin{thebibliography}{1}

\bibitem{Bhattacharjee:2009:EIC:1687553.1687573}
Bishwaranjan Bhattacharjee, Lipyeow Lim, Timothy Malkemus, George Mihaila,
  Kenneth Ross, Sherman Lau, Cathy McArthur, Zoltan Toth, and Reza Sherkat.
\newblock Efficient index compression in {DB2 LUW}.
\newblock {\em Proc. VLDB Endow.}, 2(2):1462--1473, August 2009.

\bibitem{DeanOfficialplusslides:2009:CBL:1498759.1498761}
Jeffrey Dean.
\newblock Challenges in building large-scale information retrieval systems:
  invited talk.
\newblock WSDM '09, pages 1--1, New York, NY, USA, 2009. ACM.

\bibitem{LemireBoytsov2013decoding}
Daniel Lemire and Leonid Boytsov.
\newblock Decoding billions of integers per second through vectorization.
\newblock {\em Softw. Pract. Exper.}, 45(1), 2015.

\bibitem{Stepanov:2011:SDP:2063576.2063627}
Alexander~A. Stepanov, Anil~R. Gangolli, Daniel~E. Rose, Ryan~J. Ernst, and
  Paramjit~S. Oberoi.
\newblock {SIMD}-based decoding of posting lists.
\newblock CIKM '11, pages 317--326, New York, NY, USA, 2011. ACM.

\bibitem{williams1999compressing}
Hugh~E. Williams and Justin Zobel.
\newblock Compressing integers for fast file access.
\newblock {\em Comput. J.}, 42(3):193--201, 1999.

\end{thebibliography}

\end{document}